\documentclass{andp2012}
\usepackage[english]{babel}
\title{Numerical Relativity and the Discovery of Gravitational Waves}
\author[F. Author]{Robert A. Eisenstein\inst{1,}\footnote{Corresponding author\quad E-mail:~\textsf{reisenst@mit.edu}}}
\address[1]{MIT LIGO, NW22-272,
185 Albany St.,
Cambridge, MA 02139}
\shortauthors{Robert A. Eisenstein}
\begin{abstract}
Solving Einstein's equations precisely for strong-field gravitational systems is essential to determining the full physics content of gravitational wave detections.  Without these solutions it is not possible to infer precise values for initial and final-state system parameters. Obtaining these solutions requires extensive numerical simulations, as Einstein's equations governing these systems are much too difficult to solve analytically.  These difficulties arise principally from the curved, non-linear nature of spacetime in general relativity.  Developing the numerical capabilities needed to produce reliable, efficient calculations has required a Herculean 50-year effort involving hundreds of researchers using sophisticated physical insight, algorithm development, computational technique and computers that are a billion times more capable than they were in 1964 when computations were first attempted. My purpose is to give an accessible overview for non-experts of the major developments that have made such dramatic progress possible. 
\end{abstract}
\shortabstract
\begin{document}
\maketitle
\noindent

\section{Overview of a Black-Hole Black-Hole Coalescence}\label{overview}
On September 14, 2015, at 09:50:45 UTC the two detectors of the advanced Laser Interferometer Gravitational-Wave Observatory (aLIGO)\cite{aligodescrip} simultaneously observed\cite{discovery} the binary black hole merger known as GW150914. The binary pair merged at a luminosity distance of $410^{+160}_{-180}$  Mpc.  Analysis revealed\cite{paramest} that the two BH masses involved in the coalescence were, in the source frame,  $35.8^{+5.3}_{-3.9}$ and $29.1^{+3.8}_{-4.3}$ \(M_\odot\), while the mass of the final-state BH was $62.0^{+4.1}_{-3.7}$ \(M_\odot\).  The difference in mass between the initial and final state, $3.0^{+0.5}_{-0.4}$ \(M_\odot\), was radiated away as gravitational radiation.  No associated electromagnetic radiation or other cosmic rays were observed. Astonishingly, the coalescence and ringdown to a final stable BH took less than 0.2 second (within LIGO's frequency band), coming after an orbital dance lasting billions of years.  This observation, coming 100 years after Einstein's publication of general relativity, is yet another confirmation of its validity.  It also is the first direct confirmation that BHs can come in pairs.  

\begin{figure}
 \includegraphics[width=8.5 cm]{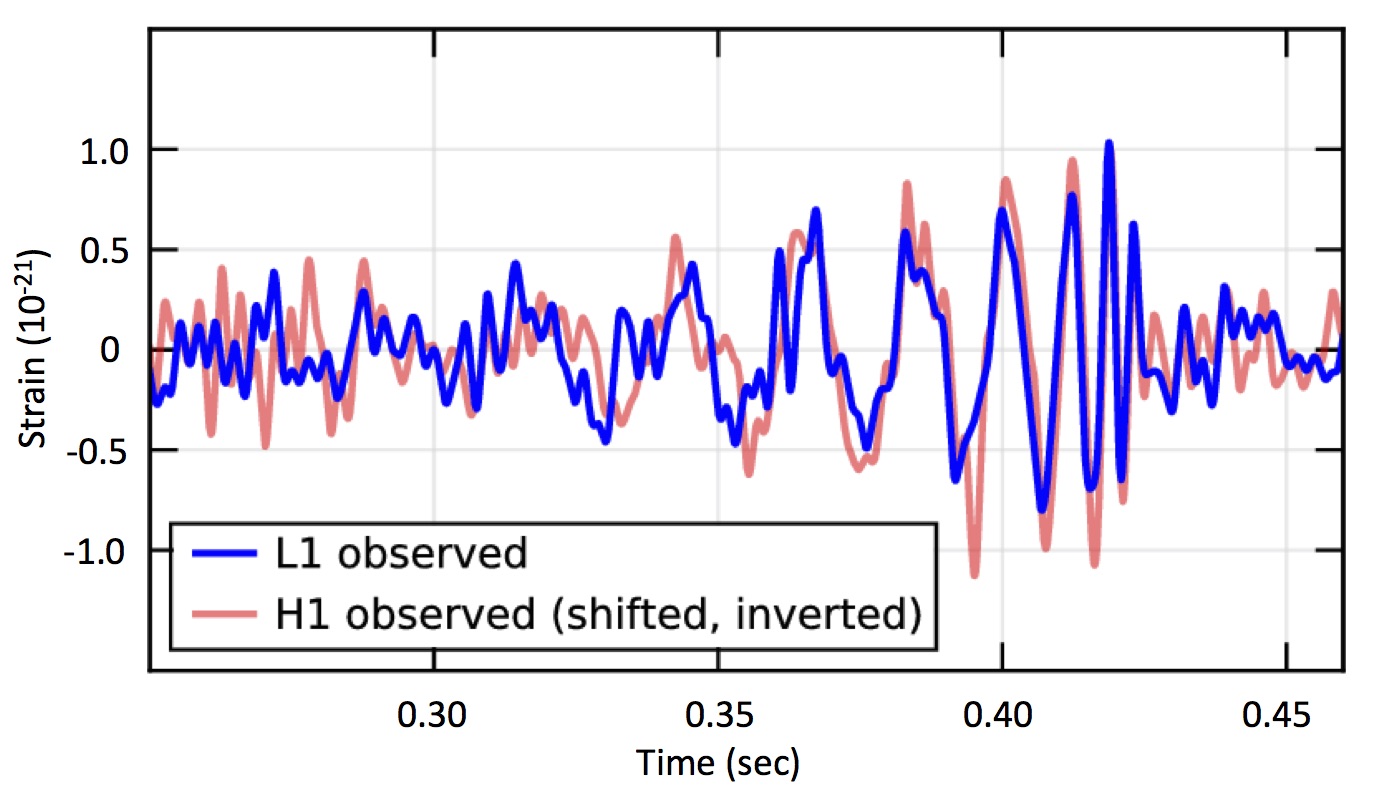}
\caption{GW strains within a 35--350 Hz passband measured at the Hanford and Livingston LIGO observatories during the detection of GW150914.  Time is measured relative to 09:50:45 UTC.  The event arrived $6.9^{+0.5}_{-0.4}$ ms later at Hanford than at Livingston (see text). 
(From Ref.\,[\citenum{discovery}]) }
\label{strainplot}\col
\end{figure}

Figure\,\ref{strainplot} is a comparison of the observed strains, within a 35-350 Hz passband, at the Hanford and Livingston LIGO sites after shifting and inverting the Hanford data to account for the difference in arrival time and the relative orientation of the detectors.  The event was identified nearly in real time using detection techniques that made minimal assumptions\cite{minassump} about the nature of the incoming wave.  Subsequent analysis used matched-filter techniques\cite{matchfil} to establish the statistical significance of the observation.   Detailed statistical analyses using Bayesian methods were used to estimate the parameters of the coalescing BH--BH system.\cite{paramest} 

\begin{figure}[h]
\centering
\includegraphics[width=8.5 cm]{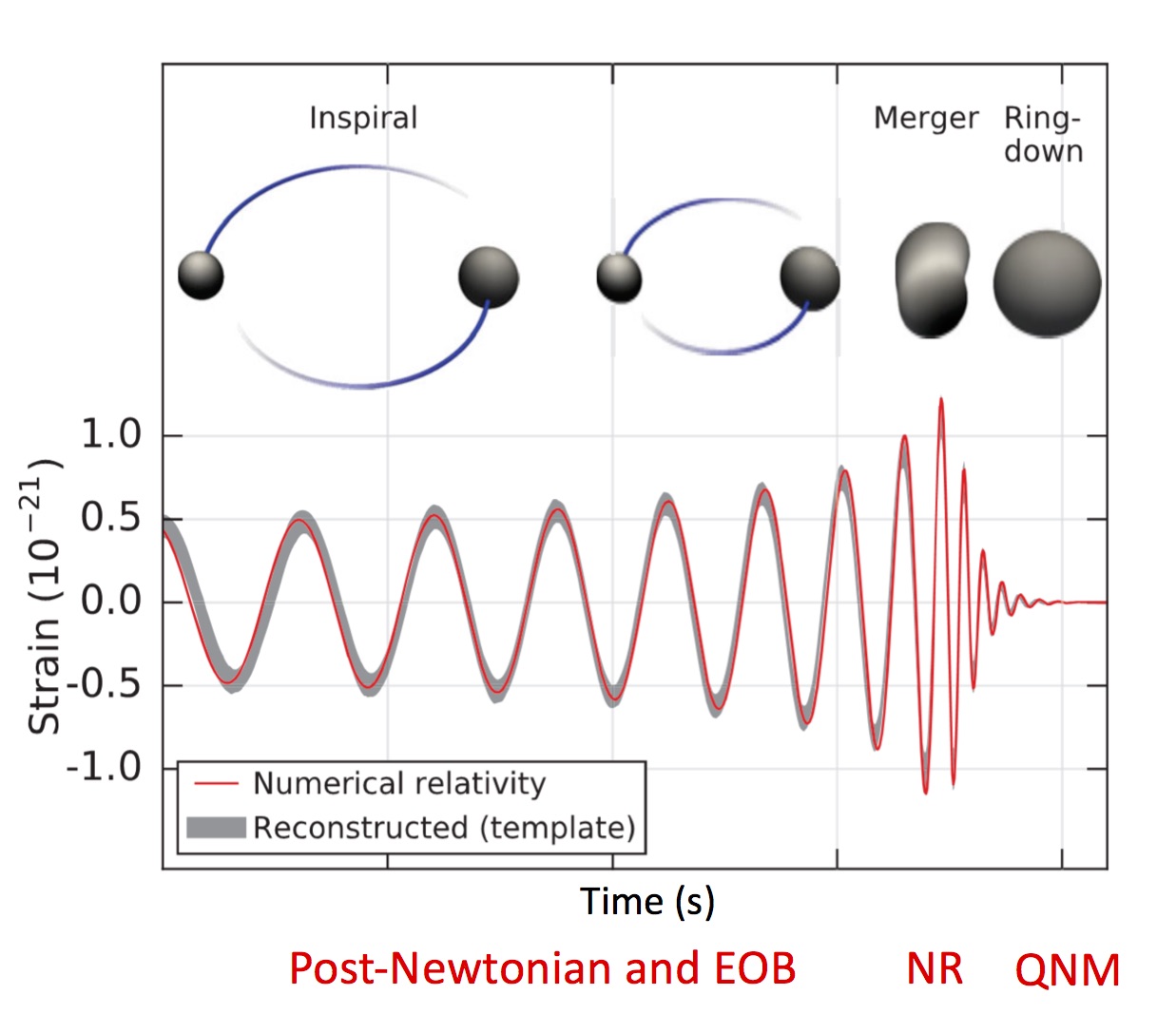}
\caption{Top: A schematic drawing of the inspiral, plunge, merger and ringdown of two coalescing BHs (see text).  Bottom: Comparison of a best-fit template of the measured strain data to the predicted unfiltered theoretical waveform, calculated using the extracted physical parameters. 
(From Ref.\,[\citenum{discovery}]) }  
\label{insp-mer-rd}
\end{figure}

Long before coalescence occurs, the two orbiting BHs can be represented as point masses co-rotating in a Newtonian orbit of very large size.  In Einstein's Universe, however, an ``inspiral" is taking place due to energy lost to gravitational radiation.This ``inspiral'' is indicated on the left side of Fig.\,\ref{insp-mer-rd}.   As it progresses, the orbit becomes circularized due to the energy loss.  The spacetime is basically flat except near each BH.  Even so, Newtonian physics cannot accurately describe what is happening.  Instead, ``Post-Newtonian'' (PN)\cite{postnewt} and ``Effective One-Body'' (EOB)\cite{eob2} methods must be employed.\nobreak
\footnote{In brief, Post-Newtonian methods utilize an expansion of Einstein equations in powers of $v/c$ to calculate the two-body BH-BH dynamics.  The results are most reliable when the gravitation is weak and internal motions are slow.  The Effective One-Body approach extends the range of standard PN by mapping the two-body problem into a single body moving in dynamics provided by an effective metric.}

As the BH's near each other (center, Fig.\,\ref{insp-mer-rd}), spacetime begins to warp and the BH horizons are distorted.
The EOB approach provides a good description (better than one might expect) until the beginning of coalescence, when the spacetime becomes significantly curved and highly non-linear.  In fact, the inspiraling waveform depends strongly on several aspects of the BH--BH interaction, \emph{e.g.}\,their masses, spins, orbit orientation and eccentricity.  This dependence plays a key role in the extraction of those parameters, but requires fits to numerical relativity simulations (Sec. \ref{imr}) to reproduce the correct result as the binary system approaches merger. Recently, parameter estimation methods have directly used numerical relativity simulations \cite{Abbott:2016apu, Lovelace:2016uwp, Williamson:2017evr} to do this.

Soon after the BH's reach their ``innermost stable circular orbit'' (ISCO) they ``plunge'' together, coalescing into a single highly vibrating, spinning (Kerr)\cite{kerrmet} BH.  Numerical relativity is needed to describe this.  The final BH rings down via the emission of gravitational radiation to a stable, spinning, non-radiating BH.  The ringdown can be described using a perturbative quasi-normal modes (QNM) model.\cite{qnm}  

An overview of the basic physics of the entire BH--BH merger is available in Ref.\,[\citenum{basicphys}]; 
a very useful discussion of what can be learned about the coalescence without recourse to numerical relativity is also provided there.  For example, the chirp mass $M = (m_1 m_2)^{3/5}/(m_1+m_2)^{1/5}$ of the binary pair can be determined to reasonable precision by using the strain waveforms (\emph{cf.} Fig.\,\ref{strainplot}) to estimate the time rate of change of the orbital frequency near coalescence.

\section{Einstein's Equations}\label{einseq}
Einstein's equations,\cite{lambourne, hartle} written in final form in November, 1915, are expressed in terms of the four generalized coordinates of spacetime, which is represented as a  geometrical \emph{Riemann manifold}\nobreak
\footnote{A Riemann manifold is a curved space which is locally flat near each spacetime point.  The Riemann curvature tensor describes it by measuring the change of a vector as it is transported around a closed path on the manifold, while always remaining parallel to its original orientation.  This is referred to as ``parallel transport.'' On a flat surface the vector will not change, while on a curved  surface it does.  Thus the Riemann tensor is identically zero for flat surfaces but not otherwise.  A Riemann manifold is defined intrinsically, without reference to any embedding of it in an exterior space.  This is discussed further in Sec.\,\ref{admpr}.} 
$\mathcal{M}$ that, unless compacted, extends to infinity in all directions.\cite{dennis}    At this stage, they are not represented by a specific coordinate system.  The manifold shape is determined by the real 4-by-4 metric tensor $g_{\mu\nu}$, which in Einstein's theory is determined by the mass densities and energy fluxes present at every point in spacetime.  In 4-space we use Greek letters for the indices; three of the coordinates (labeled 1-3)  are spatial and one (labeled 0) represents time.These relationships are summarized by Einstein's equations written in tensor form:\cite{einsteq1}\nobreak
\footnote{Einstein's equations are often written using units in which the speed of light ($c$) and Newton's gravitational constant ($G$) are set equal to 1.  Thus one \(M_\odot\) $\simeq 1.5km \simeq 5{\mu}s$.}

\begin{equation}\label{eq:1}
G_{\mu\nu} := R_{\mu\nu} - \tfrac{1}{2} \, g_{\mu\nu}R = 8{\pi}T_{\mu\nu}
\end{equation}                        
The quantity $G_{\mu\nu}$, Einstein's tensor,\nobreak
\footnote{The Einstein tensor measures the curvature of the manifold in a region near each point.}
is defined in terms of the metric tensor $g_{\mu\nu}$, the Ricci curvature tensor\nobreak
\footnote{The Ricci tensor measures the difference in geometry between a Riemann metric and ordinary Euclidean $n$--space.}
$R_{\mu\nu}$ and the Ricci scalar\nobreak
\footnote{The Ricci scalar is a real number that measures the intrinsic geometry of a Riemann manifold near a given point.}
$R = \gamma_{\mu\nu} R^{\mu\nu}$ (using the Einstein convention to sum over repeated indices).  The energy-momentum, or stress-energy, tensor is represented by $T_{\mu\nu}$.

At first glance it appears that in Einstein's equation the geometry of spacetime appears only on the left-hand side, imbedded in $G_{\mu\nu}$, while the physical momentum-energy content appears only on the right, imbedded in $T_{\mu\nu}$.  In fact this is generally only true for Einstein's equations in vacuum.  Otherwise, the metric $g_{\mu\nu}$ can also appear in the expressions for the stress-energy tensor.  Nonetheless, as John Wheeler memorably remarked\cite{jawheeler}, Eq.\,\ref{eq:1}  shows that ``Matter tells spacetime how to curve, and spacetime tells matter how to move.''

The metric tensor $g_{\mu\nu}$ plays the same role in general relativity as it does in special relativity.  In each case it provides the link between the generalized coordinates $x_{\mu}$ and the invariant spacetime interval $ds$:  ${ds}^2 = g_{\mu\nu}  {dx}^{\mu} {dx}^{\nu}$.  In special relativity it defines a flat (Minkowski)\nobreak
\footnote{Minkowski space is described by a flat 4-dimensional manifold in which the time coordinate is treated differently than the three space coordinates.  Thus Minkowski space, though flat, is not a 4-dimensional Euclidean space.}
space.  In general relativity it defines the curved (Riemann) manifold $\mathcal{M}$.  The curvature, due to gravitational sources, enters via the Ricci tensor $R_{\mu\nu}$ and the Ricci scalar R.    Thus in both special and general relativity, the metric tensor elements determine all the spacetime observables we can calculate.  

The subscripts $(\mu,\nu)$ range over the integers 0 to 3, implying the need to solve a system of 16 coupled equations.  However, the symmetries of the metric $(g_{\mu\nu} = g_{\nu\mu}, \mu \neq \nu)$ limit the actual number to 10.  The simple appearance of Einstein's equations in tensor notation masks a very great deal of complexity.  When written out in full they can contain thousands of terms.  These will have significant non-linearities due to the spacetime curvature that occurs when the gravitational fields are strong.

\section{Solving Einstein's Equations}\label{soleeq}
Due to the complexities mentioned above, there are very few analytical solutions of Einstein's equations of physical relevance.  The ones we know of arise in situations involving a high degree of symmetry.  Most important for the present discussion are the Schwarzschild solution\cite{schwmet}  (for a spherically-symmetric mass M with spin 0) and the Kerr solution\cite{kerrmet} (for a spherically-symmetric mass M with spin J).  Exact solutions that include a charge Q on the BH (an unlikely prospect) have also been found\cite{rnmet, knmet} but will not be discussed here.  

Schwarzschild's 1916 discovery led to one of the most important predictions of general relativity: the existence of BH's.  A valuable simplification comes in the form of the ``no-hair'' conjecture,\cite{nohair} which states that in four dimensions the solutions to Einstein's equations for a stationary BH can only depend on its mass, spin and charge.  

Einstein predicted the existence of gravitational waves\cite{flanhughes}\nobreak
\footnote{Gravitational waves are ripples in spacetime itself rather than a disturbance superimposed on it (\emph{e.g.} emission of an electromagnetic wave from a vibrating charge).  Although gravitational waves carry energy and can do work, they are absorbed only very weakly and so can travel cosmological distances at speed $c$ without dispersion.  This has been confirmed recently\cite{speedofgw} to about 1 part in $10^{15}$}
moving at the speed of light\cite{speedofgw} in 1916.  Reasoning by analogy to electromagnetism (\emph{i.e.}\,accelerating masses should radiate gravitational waves as accelerating charges radiate electromagnetic ones),\nobreak
\footnote{An essential difference is that the lowest order of electromagnetic radiation is the dipole term, while for gravitational radiation it is the quadrupole.  So any source of gravitational waves must possess mass distributions with time-varying quadrupole and/or higher multipole moments.}
he found them by linearizing the Einstein equations for the case of nearly flat spacetime (\emph{i.e.}\,weak gravitation).  In the strong--field case, where the field equations are not linear and spacetime is itself evolving, and not flat, the definition of what constitutes a wave is less clear.  What does it mean to separate the wave from the spacetime?

Nonetheless, it would seem that gravitational waves must exist in that situation also.  But for many years there was considerable uncertainty as to their existence, even from Einstein himself, but the issue was put to rest\cite{stickybead, sormani, kennefick} in the mid-1950's.    Finally, it is the full non-linear equations that must be solved numerically in order to quantify the nature of BH--BH, BH--Neutron Star (NS) or NS--NS coalescences.  We return to this discussion in Sec.\,\ref{gravwave}.
 
As if BHs and gravitational waves were not enough, Einstein's equations also predict that the structure of the Universe is not static: as time goes on, it will either expand  or contract. Since there was no evidence in 1916 for either of these prospects, Einstein introduced a ``cosmological constant'' $(\Lambda$) to force his equations to predict a static Universe.  When the expansion of the Universe\cite{netab} was established in 1929, he later called this decision ``my greatest blunder.''  Ironically, with the discovery\cite{acceluniv} in 1998 that the Universe is accelerating as it expands, the cosmological constant plays an important role in accounting for (if not understanding) the cosmic acceleration.  Today, Einstein's equations with the cosmological constant included, form the basis of the ubiquitous Friedmann-LeMa\^itre-Robertson-Walker ``standard model" for a homogeneous, isotropic Universe.\cite{lambourne2, ryden}

\section{Numerical Relativity and BH--BH Coalescence\cite{hawke, baumshap1, alcubierre1, gour1, centrev}}\label{nrbhc}
It is worth pointing out that even though these calculations are prodigiously difficult, the BH--BH system -- because it contains only gravitational fields and no matter distributions -- is very likely the simplest strongly--interacting gravitational problem we will ever encounter.  If the study of strong-field general relativity is to have a future, it is imperative to solve it.  In addition, the BH--BH coalescence problem is particularly important because it affords a clean test of our understanding of strong--field general relativity.

The long road to achieving stable, accurate
numerical solutions began in 1952, when Four\`{e}s-Brouhat\cite{ycb1} showed that Einstein's vacuum equations were, locally at least, \emph{well-posed}.  Simply put, this means: (1) that solutions of the equations exist; and (2) that small changes to initial conditions produce only stable, continuous (\emph{i.e.}\,non-chaotic) changes in the output.  Given the difficulty of Einstein's equations, these seemingly reasonable expectations are far from obvious.  As an example, see the work of Choptuik\cite{critphenom} on the appearance of critical phenomena in general relativity.

Four\`{e}s-Brouhat's proof that Einstein's vacuum equations could be posed, at least locally, as a unique, stable initial value problem was based on certain smoothness assumptions\cite{mholst} and the use of \emph{harmonic coordinates}\nobreak
\footnote{\,\,The harmonic coordinates $x^\nu$ are defined via a wave equation:  $\nabla_{\mu}{\nabla}^{\mu} x^{\nu} = 0$ for ${\nu}=0,1,2,3$.}
to specify the evolving spacetime.\cite{friedrich1, garf, mholst}  This procedure turns Einstein's equations into a set of ten quasi-linear wave-like equations with favorable (\emph{hyperbolic}) 
stabiliity properties (see Sec.\,\ref{evcon}).

\subsection{The ADM Procedure}\label{admpr}
During the next several decades, many substantial difficulties\cite{vitor, cardoso} had to be overcome to obtain stable, accurate solutions. The first was to recast Einstein's equations in the form of a computable, time-step iteration process (\emph{i.e.} an initial value problem) that would evolve from initial conditions (\emph{i.e}.\,an initial spacetime), through BH--BH coalescence, to the final state.   In the world of partial differential equations (PDE's) this is called a \emph{Cauchy problem.}  In general relativity, this recipe is referred to as a ``3+1'' approach because space and time are separated.  This formulation comes at an ironic price: giving up overall covariance after Einstein worked so hard to incorporate it into General Relativity.   It was first proposed by Arnowitt, Deser and Misner\cite{adm} (ADM) in 1962.  

In 1979, York rewrote\cite{york} the original ADM prescription to emphasize its role in evolving the Einstein equations\cite{alcubierre2} rather than as a basis for a theory of quantum gravity (the original intent of the ADM work).  His treatment is now ubiquitously referred to as the ADM prescription.  It has spawned many close cousins, all of which are referred to as 3+1 algorithms. 

\begin{figure}[h]
\centering
\includegraphics[width=8.0 cm]{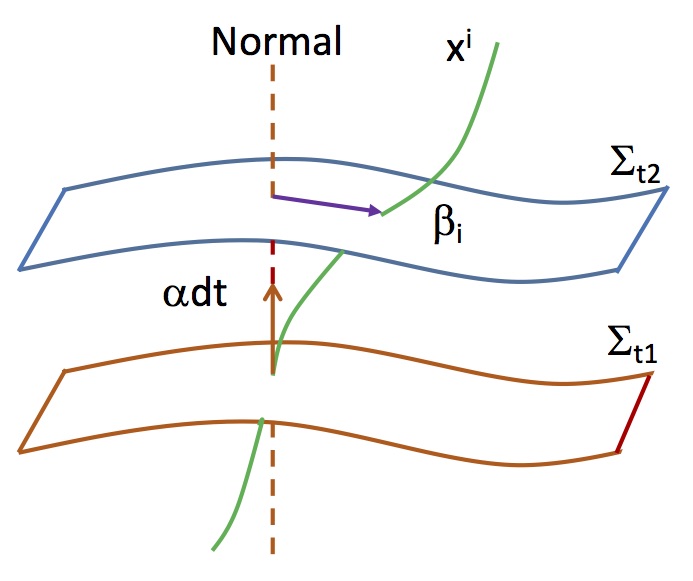}
\caption{A schematic 3+1 ADM decomposition.  $\Sigma_{t1}$ and $\Sigma_{t2}$  are spacelike 3-dimensional foliations separated by coordinate time $t_2 - t_1$.  The quantity $\alpha\,dt$, with $\alpha$ the lapse, is the proper time step between  $\Sigma_{t2}$ and $\Sigma_{t1}$. The shift $\beta_i$ measures the change in coordinate $x^i$ in moving from the earlier foliation. The green line is the origin of the spatial coordinate system.}
\label{bob3+1}
\end{figure}

The basic ADM idea is to decompose the spacetime by creating a stack of 3-dimensional, spacelike ``foliations'', or slices, each characterized by a fixed coordinate time (see Fig.\,\ref{bob3+1}).  These we label $\Sigma_t$.  The system evolves by moving with time from one foliation to the next.  The invariant spacetime interval, ${ds}^2 = g_{\mu\nu}  {dx}^{\mu} {dx}^{\nu}$ in 4-space, becomes in the 3+1 description:

\begin{equation}\label{spacetime}
ds^2 = -{\alpha}^2 dt^2 + {\gamma}_{ij} ( dx^i + {\beta}^{i} dt ) ( dx^j + {\beta}^{j}dt) 
\end{equation}

Here the $\gamma_{ij}$ are the 3-dimensional metrics for these spacelike surfaces, labeled with Latin indices \emph{i} and \emph{j} running from 1 to 3.  Note that time appears explicitly.  The quantity $\alpha$ (the \emph{lapse}) and the three $\beta_{i}$ (the \emph{shift vector} \vec{\beta}) are \emph{gauge variables}\nobreak
\footnote{\,\,Gauge freedom allows modification of physical equations to improve solubility as long as the physical observables (here the $g_{\mu\nu}$) don't change.  Since these cannot depend on the coordinate system, changing it is an example of gauge freedom.  Another is the use of scalar and vector potential functions $\phi$ and $\bf{A}$ in Maxwell's equations (Ref.\,[\citenum{baumshap1}], Ch 11, p. 378).}
that may be freely specified but must be chosen with care.  Effectively they represent the coordinate freedom inherent in Einstein's equations.  

The lapse determines the rate in proper time at which one progresses from one slice to the next, while the shift vector  basically quantifies how much the spatial coordinates change between foliations.  Both are usually allowed to vary dynamically as the geometry of the system evolves; effectively, we are choosing coordinates as we proceed.  

There are a number of standard prescriptions for choosing the lapse.\cite{lapseshift1}  Good choices will avoid singularities, improve convergence and speed up the calculation.  One finds frequent references to ``maximal slicing," a choice that avoids singularities, is simple mathematically but computationally expensive.  Variations include the ``Bona-Masso" and ``harmonic slicing" families.  Yet another choice is the ``$1+log$" family, which avoids singularities in a way similar to maximal slicing but is more economical to implement.  A choice to avoid is the so-called ``geodesic slicing," where $\alpha$ is set equal to 1 so that the foliations are equally spaced in time.  In that case observers moving with the foliations are in free fall.  A more detailed examination shows that this almost always leads to a singularity.

Regarding the shift vector, the standard choice in recent work studying BHs is the so-called ``$\Gamma$-driver" condition.\cite{lapseshift1}  Its purpose is to constrain the large field gradients that can appear near a BH.  

Because the foliations $\Sigma_t$ are embedded in the overall spacetime manifold $\mathcal{M}$, they are characterized by the real \emph{Extrinsic Curvature Tensor} $K_{ij}$ that describes the nature of the embedding.\nobreak
\footnote{$\mathcal{K}_{ij}$ measures the change in direction of a surface normal vector under parallel transport (Ref. [\citenum{alcubierre1}], Ch. 2, p. 69). }
The purpose of this tensor is to separate the intrinsic curvature of the foliation (\emph{i.e.} the $\gamma_{ij}$) from the extrinsic curvature due to the way it is embedded in the overall spacetime.
Eq.\,(\ref{defK}) can be considered a definition of $K_{ij}$.  Note its relation to the time derivative of $\gamma_{ij}$.


Despite the promise of the ADM method, evolving a BH--BH system through coalescence remained elusive. The reason was that its equations were shown to be only weakly hyperbolic\cite{illposed, pde} and thus are ill-posed.  We next discuss the resolution of this issue.

\subsection{ADM Evolution and Constraints} \label{evcon}
As mentioned in Sec.\,\ref{einseq}, the symmetries of the metric tensor reduce Einstein's set of 16 equations for the $g_{\mu\nu}$ to 10 coupled, non-linear PDE's.  Six of these contain space and time derivatives up to first order in time and second order in space.  These equations provide the \emph{evolution} of the spacetime.  The remaining four equations, containing no time derivatives, are \emph{constraint} equations.

Since our focus in this paper is mostly on BH--BH coalescence, in what follows we restrict the discussion to the vacuum case (\emph{i.e.} the absence of matter or energy sources).

\textbf{Evolution equations.}  
In the ADM procedure there are three evolution equations for the spatial metric $\gamma_{ij}$:
\begin{equation}\label{defK}
{\partial}_{t}\gamma_{ij}=-2\alpha K_{ij} + D_{i}\beta_{j} + D_{j}\beta_{i},
\end{equation}
and three for the extrinsic curvature $K_{ij}$:
\begin{equation}
 \begin{split}
{\partial}_{t} K_{ij} = & \,\alpha (R_{ij} -  2 K_{ik} K^{k}_{j}   + K K_{ij} ) - D_{i} D_{j}\alpha +  \\
& \beta^{k}\partial_{k} K_{ij} + K_{ik}\partial_{j}\beta^{k} + K_{kj}\partial_{i}\beta^{k}.
\end{split}
\end{equation}
These six equations\cite{evconeqs} provide the evolution of the initial spacetime.  
Here the symbol $\partial_t $ is an ordinary partial derivative with respect to time, $D_i$ is a spatial covariant derivative and $K$ is the trace of $K_{ij}$: $K=\gamma_{ij}K^{ij}$.  These equations contain mixtures of \emph{hyperbolic} and \emph{parabolic}  (\emph{i.e.}\,time-dependent) behavior.  

Hyperbolic equations are basically wave equations that describe wave propagation at finite speed.\cite{pde}  Solutions to wave equations are generally very stable and converge rapidly.  On the other hand, real parabolic equations (\emph{e.g.}\,the heat equation) do not exhibit wave-like behavior.  However, a parabolic equation with an imaginary component (\emph{e.g.} the Schr\"{o}dinger equation), exhibits both a wave speed and dispersion.\nobreak
\footnote{\,\,I thank F. Tabakin for pointing this out.}
 
\textbf{Constraint equations.} 
The remaining four equations contain no time derivatives. They are referred to as constraint equations because their solutions, while evolving with time, must do so in a way so that the constraint equations are always satisfied.  They are also independent of the lapse $\alpha$ and shift $\vec{\beta}$. 
One equation is referred to as the \emph{Hamiltonian constraint}:
\begin{equation}
^{(3)}R+K^2-K_{ij}K^{ij}=0,
\end{equation}
and the remaining three are called the \emph{momentum constraints}:
\begin{equation}
D_{j}(K^{ij}-\gamma^{ij}K)=0.
\end{equation}
Here $^{(3)}R$ is the trace of the spatial Ricci tensor $R_{ij}$.
These are \emph{elliptic} (\emph{i.e.}\,time-independent) equations.  Elliptic equations are often used to describe time-independent boundary-value problems.  Because of the non-linearity of strong-field general relativity, they are harder than usual to solve numerically. 

As stated above, once the constraints are satisfied initially, mathematically they will remain that way.  But for numerical solutions that is often not the case, especially when significant non-linearities are present.  Small numerical errors can exponentially grow.  Keeping the constraints satisfied at all times has proven essential to reaching stable, convergent solutions of the BH--BH coalescence problem.

An instructive parallel appears in Maxwell's equations.  There, the laws of Amp\`{e}re and Faraday, both containing time derivatives of the electric and magnetic fields \textbf{E}  and \textbf{B}, are the evolution equations, while Gauss's Laws for \textbf{E} and \textbf{B} serve as constraints.  Since these equations are linear the constraints are usually well-behaved.  When they aren't, the results are not solutions to Maxwell's equations.  The analogue is true in numerical relativity.

For greatest stability, the evolution equations should be as wave-like (hyperbolic) as possible.  Gauge freedom is useful for this purpose, keeping in mind that poor gauge choices (including non-optimal coordinates\cite{stupid}) can adversely affect well-posedness.  The constraint equations have proven very useful here.  Since they can always be written in the form $\mathcal{C}(x, y, z) =0$  (\emph{e.g.}\, $\bf{\nabla \cdot E}$ -- $4 \pi \rho = 0)$, one can add them (or multiples of them) to the evolution equations wherever that might be useful.  One can also check for constraint violations by evaluating $\mathcal{C}(x, y, z)$ as the numerical evolution proceeds.  

There are many other ways\cite{lapseshift1}
to use gauge freedom to control problems arising from stability and convergence issues, physical or coordinate singularities, numerical round-off error, and issues associated with boundary problems at BH horizons (among others).  Perhaps the most important lesson in the development of numerical relativity is that gauge choices (including the choice of coordinates) are every bit as important as computing power.

The 1987 work of Nakamura, Oohara and Kajima, presented\cite{nok} a version of ADM that showed much better stability.  Later, Shibata and Nakamura\cite{shibnak} (1995) and Baumgarte and Shapiro\cite{baumshap2} (1998) confirmed and extended those results.  These efforts are commonly known as the \emph{BSSNOK} approach. It was essential to achieving full 3-dimensional simulations of BH--BH coalescences and is in wide use today.  It confirms the importance of selecting carefully the best formulation of Einstein's equations for the problem at hand.

To orient ourselves, let us count the number of degrees of freedom in the ADM prescription.\cite{sperhake} The overall metric $g_{\mu\nu}$, Eq. (\ref{spacetime}), depends on the six  ${\gamma}_{ij}$ as well as the lapse and the shift (which however are free gauge variables and do not contain physical information).  The four constraint equations impose conditions on the ${\gamma}_{ij}$ at each foliation as the evolution proceeds.  This leaves two ``gravitational" degrees of freedom which we associate with the $+$ and $\times$ gravitational wave polarization modes that will discussed in Sec. [\ref{gravwave}].  

\subsection{Generalized Harmonic Coordinates with Constraint Damping (GHCD)}\label{harmcon}
Beginning with Einstein, harmonic coordinates have played a major role in general relativity.\cite{ghhistory}  As noted in Sec.\,\ref{nrbhc}, they were used by Four\`{e}s-Brouhat\cite{ycb1} to show the local well-posedness of Einstein's vacuum equations. This occurs because they convert the equations into a second-order strongly hyperbolic form.  

Today, in a generalized form, harmonic coordinates are important in solving numerically the BBH coalescence problem.  Based on earlier work by Friedrich,\cite{friedrich2} Garfinkle,\cite{garf} and Szil\'agyi and Winicour,\cite{szilagyi} Pretorius\cite{pretorius, pretorius2} used them to describe a completely different means of achieving a stable numerical coalescence of a BH--BH pair.  An extended treatment was soon provided by Lindblom \emph{et al.}\cite{lindblom}

The generalization takes the form of added \emph{source terms} $H^\nu$ to the original wave equations\footnote{\,See footnote 10.}  
for the harmonic coordinates, \emph{viz.}: $\nabla_{\mu}{\nabla}^{\mu} x^{\nu} = H^\nu$ for ${\nu}=0,1,2,3$.  These source terms are taken as independent functions.  They allow the introduction of arbitrary gauge conditions in a manner analogous to the use of the lapse and shift in the ADM prescription.  

Another major departure is that in the original GHCD method\cite{pretorius, pretorius2} the (second order) metric itself is directly discretized.  Constraint damping terms are added to achieve a stable evolution of the spacetime.\cite{pretorius, gundlach}  The work of Lindblom \emph{et al.} is a generalized harmonic evolution development that is fully first order in time and space. In the ADM treatment one is dealing with equations for the $\gamma_{ij}$ and $K_{ij}$ that are first-order in time and up to second order in space.  

\subsection{Initial Conditions}\label{initconditions}
To simulate accurately a binary collision between moving, spinning BH's, we must supply initial system data, solve the constraint equations, and then maintain the constraints throughout the evolution.  

In the BSSNOK approach, the initial data consist of entries for the $\gamma_{ij}$ metric and the curvature matrix $\mathcal{K}_{ij}$, 12 real numbers in all,    thus representing a system with 12 degrees of freedom.  However, these cannot all be chosen independently because of the four constraint equations.  While they certainly should depend on the initial parameters of the BH's, and the orbital dynamics, there is no obvious relationship linking them.  So it is not clear which eight of the data to choose as free parameters and which four to use in solving the constraint equations.  

This difficult problem has been studied extensively.\cite{pfeiffer, cook, baumshap7, alcubierre4, lichnerowicz, york, brilind, bowyork, branbrug, york2, thinsand}  Until recently, the most common approach (``conformal decomposition") was that developed by Lichnerowicz\cite{lichnerowicz} and extended by York,\cite{york} who found a means of breaking the problem into smaller pieces in order to solve the four coupled constraint equations.  Building on an earlier (1963) solution by Brill and Lindquist\cite{brilind} for $N$ black holes momentarily at rest, Bowen and York\cite{bowyork} produced (in 1980) a solution for multiple black holes with arbitrary linear and angular momentum.  In 1997, Brandt and Br\"ugmann\cite{branbrug} generalized the Brill-Lindquist ansatz to a more convenient topology for the black holes.  It can also accommodate arbitrary linear and angular momentum.  Subsequently Pfeiffer and York developed (in 2005) the ``conformal thin sandwich\cite{york2, thinsand} approach; it is capable of handling a larger range of spin variables and is in wide use today.

In the original GHCD treatment the initial data were created in a much different way.\cite{pretorius}  There, two moving unstable scalar field profiles are established, with initial amplitudes, separations and boosts chosen to approximate the orbit of the co-rotating BH pair.  The BH's form when the scalar fields collapse.

Of course a crucial part of setting up the initial conditions is to choose a means for handling the physical singularities of the BH's themselves.  We turn our attention to that next.

\subsection{Excisions, Moving Punctures and Trumpets}\label{exmp}
\begin{figure}[h]
\centering
\includegraphics[width=8.0 cm]{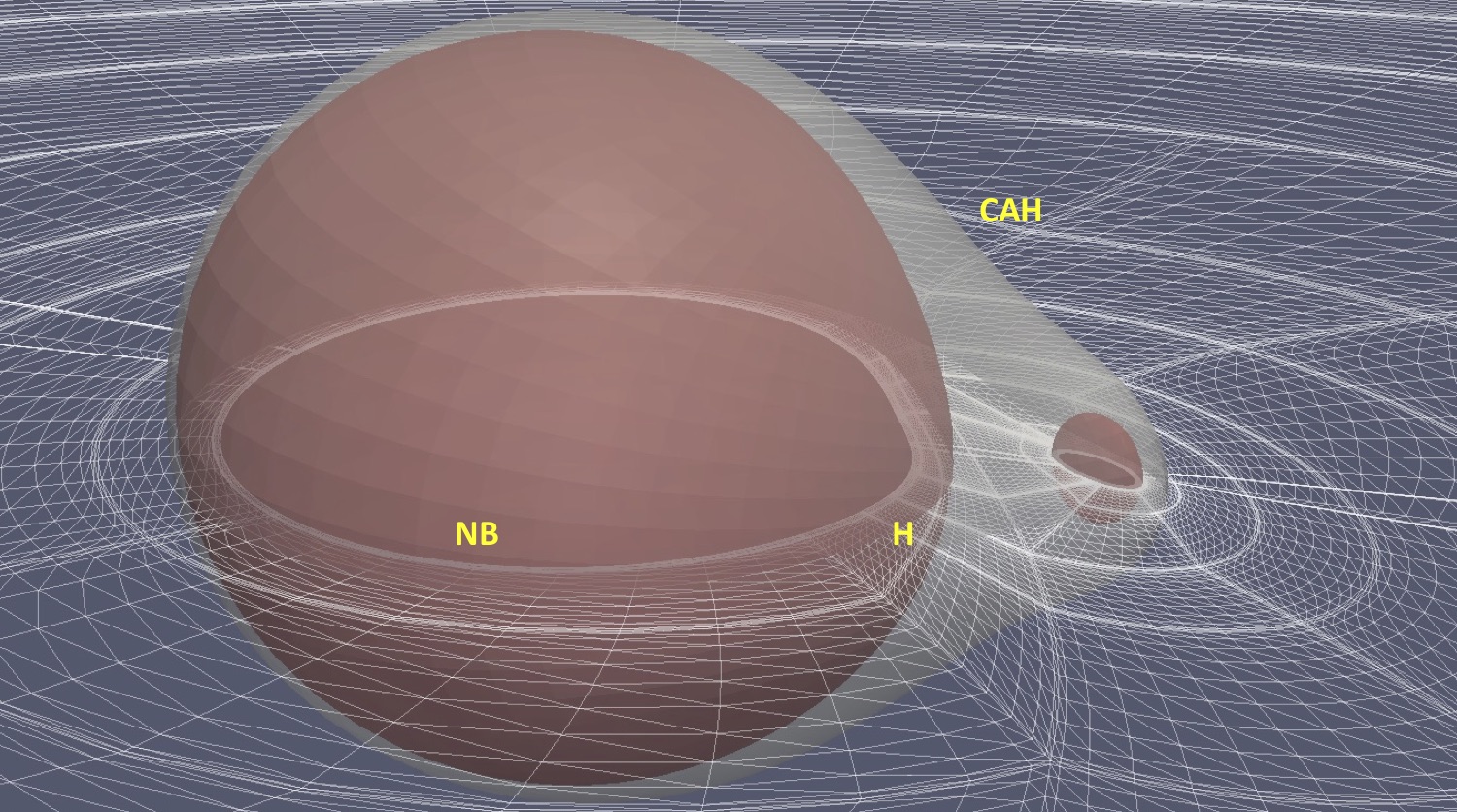}
\caption{A body-shaped, two-center coordinate system for unequal mass BHs.   ``H'' labels a BH horizon while ``NB'' is its numerical boundary.  No mesh is needed beneath that surface. ``CAH'' is the common apparent horizon.    At far distances the coordinate lines are close to spherically symmetric. (From M. Scheel, used with permission)}
\label{scheel}
\end{figure}

We are dealing with simple Schwarzschild or Kerr BHs, having event horizons behind which the singularites are hidden and out of reach physically (an idea known as \emph{cosmic censorship}\cite{coscens}).   It led William Unruh\cite{excision} to suggest in 1984 that BH singularities could be \emph{excised} from the calculation so that their influence is never felt outside the horizon. Thus information can flow into, but not out of, a BH.  Fig.\,\ref{scheel} shows the imminent coalescence of two non-equal BH's viewed from this perspective.  Note the numerical boundary just inside the BH horizon.

However, excision comes at the expense of very demanding boundary conditions.\cite{szilagi}  The BH horizons are usually of an irregular shape which is changing dynamically and is in continuous motion.  Thus spurious numerical artifacts can arise (including the unphysical emission of gravitational radiation), making fine-tuning the calculation and enforcement of the constraints a continuous necessity.  

\begin{figure}[h]
\centering
\includegraphics[width=8.0 cm]{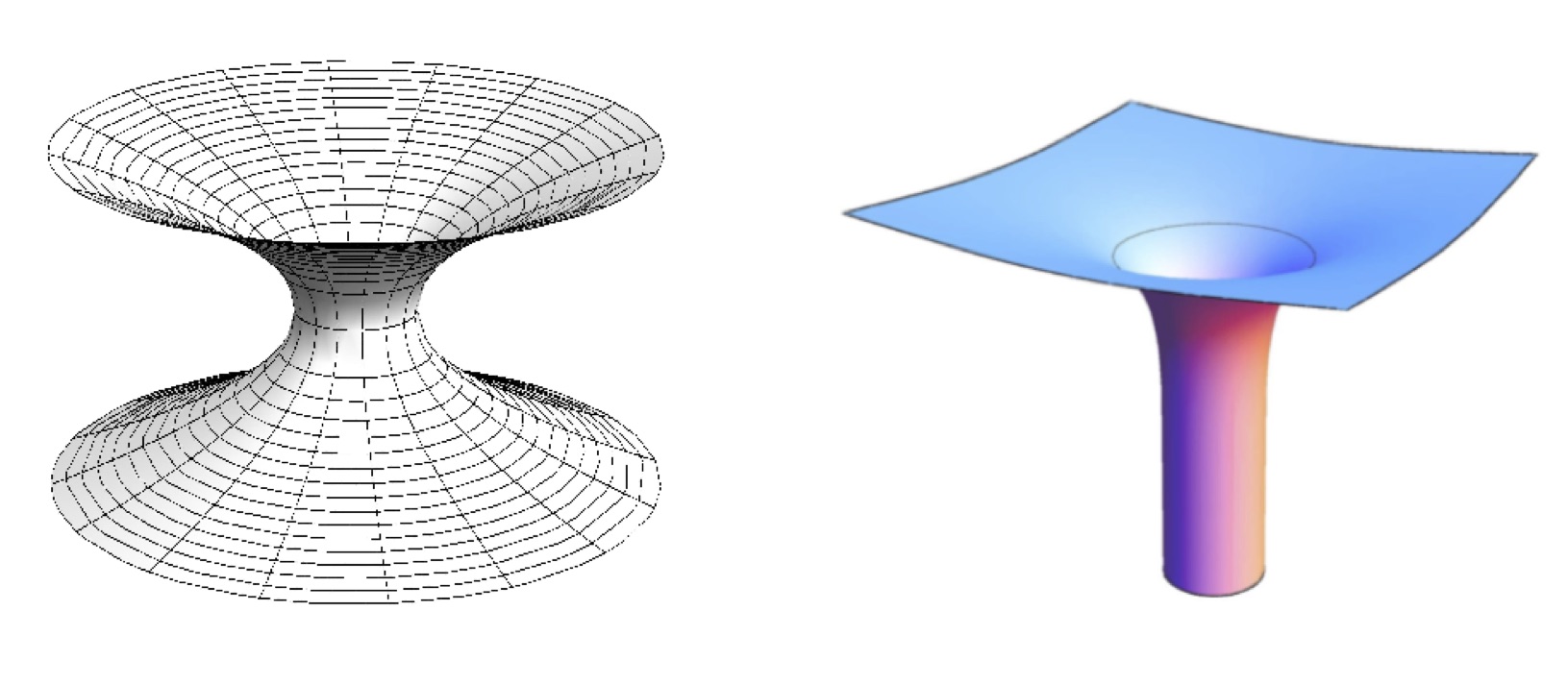}
\caption{Wormhole (left) and trumpet (right) representation of a BH.
(From Refs.\,[\citenum{hannam}], [\citenum{denn}]) }
\label{erb+trumpet}
\end{figure}

Another approach is to view BHs\nobreak
\footnote{\,\,The term ``Black Hole'' was coined by John Wheeler in 1967.}
as Einstein-Rosen bridges\cite{bridges} or ``wormholes" (See Fig.\,\ref{erb+trumpet}).  This was done first by Hahn and Lindquist\cite{hahnlind} in their seminal 1964 calculations of an axisymmetric BH--BH coalescence that founded numerical relativity. The physical singularity for the BH lies on the wormhole axis perpendicular to the spacetimes that are above and below.  Note that the coordinate lines can approach the singularity but cannot reach it.  Hahn and Lindquist modeled their (equal mass) BH--BH system as a manifold with two such Einstein--Rosen bridges opening from the same flat 3-space, but whose mouths are joined together ``below" without intersecting other spacetimes. Pictorially it resembles a jug handle (\emph{cf.} Figs. 2 and 3 in Ref.\,\,[\citenum{hahnlind}]).  The BHs must have equal mass for the throats to join together smoothly. This model was invented by Wheeler.\cite{wheelerjug}

In further developments the wormholes were compactified into \emph{punctures} (singularities) in the spacetime manifold.  This was done by factoring the spatial metric into an analytic part that represents the BH singularity (the puncture) and a second (background) part that could be evolved numerically.  In the original treatments\cite{brilind, bowyork, branbrug} the BH punctures were placed at fixed spatial coordinates.

However, this treatment led to numerical instabilities due to strong field gradients as the BH's converged. Other pathologies also arose as the coordinate landscape twisted in response to the system evolution while the singular parts remained at fixed positions.  But in 2005, almost by accident, it was discovered that a slight modification in the $``1+log"$ and ``$\Gamma$--driver" gauge conditions produced a stable algorithm (\emph{moving punctures\cite{brownsville, goddard}}) that allowed the BH's to move and finally to merge.  In this approach the singularity is not factored out from the smooth background piece but instead moves freely as the spatial metric changes.  Care is taken to ensure that the numerical calculations avoid both physical and coordinate singularities.  Allowing the punctures to move was the last step in bringing the BSSNOK evolution to a physically relevant convergence.

Subsequently it was realized that as the evolution proceeds, the initial puncture data soon evolve into a \emph{trumpet} topology\cite{hannam, denn} (see Fig.\,\ref{erb+trumpet}) that could be identified\cite{mpasbh} as a moving BH.  To improve accuracy, Hannam \emph{et al.}\cite{bowyorktr} extended the Bowen-York\cite{bowyork} initial data prescription to include a trumpet structure from the onset of evolution.

\subsection{Meshes, Coordinates, Numerical Integration}\label{mesh}
\vspace{-3mm}
The spatial extent of a BH--BH coalescence is huge.  At the beginning, the BHs are widely separated and spacetime is essentially flat except near the BH horizons.  Post-Newtonian physics holds sway.  Just before coalescence, the BH's are only tens to hundreds of kilometers apart, spacetime is highly curved and general relativity is dominant. Solving this problem involves very different length scales as it moves toward coalescence, with corresponding changes required in the numerical meshes.  \emph{Adaptive Mesh Refinement} schemes \cite{amr} have been developed to handle this issue.

The same consideration applies to the choice of coordinate system.  For a BH--BH system, it is natural to choose one that has two spherical-polar centers in close, evolving into nearly spherical symmetry far away (see Fig.\,\ref{scheel}).  In addition, much better numerical accuracy in satisfying the boundary conditions at the BH horizons will result if the coordinate lines are perpendicular to the BH horizon surface.  We must also account for the motion of the BHs and the distortion of their horizons as the coalescence evolves.  As Fig.\,\ref{scheel} shows, this can lead to great numerical complexity and the clear need to use curvilinear coordinates and non-rectangular mesh schemes.

The numerical integration procedures in most common use are finite difference (FD)\cite{choptuik} or spectral interpolation (SpEC)\cite{kidder1, kidder2, boyd, senr1} methods.  Both have long, well-known histories.  FD methods yield approximate solutions to PDEs at specific points on the mesh.   Spectral methods utilize smooth functions fitted to several mesh points that can provide highly accurate values at any location.

\subsection{Extraction of Gravitational Waves}\label{gravwave}
Gravitational waves from a BH--BH coalescence are ripples in the fabric of spacetime.  They will be reflected in the metric $g_{\mu\nu}$, which contains all the information we can learn about the coalescence.  We obtain it by analyzing the outgoing gravitational wave strains appearing in our detectors.  

If the coalescence did not disturb the spacetime too greatly, we could rely on a linearized version of the Einstein equations and write the metric $g_{\mu\nu}$ as as a small perturbative piece $h_{\mu\nu}$ added to a Minkowski spacetime background term $\eta_{\mu\nu}$:
\begin{equation}
g_{\mu\nu} = h_{\mu\nu} + \eta_{\mu\nu}, \:\:\:\:\:  \parallel h_{\mu\nu}<<1 \parallel.
\end{equation}
This is how Einstein discovered gravitational waves in 1916.  This linear approximation also led him to his famous quadrupole formula for the radiative energy loss via gravitational radiation in that limiting case:
\vspace{1mm}
\begin{equation}
L_{GW} = - \frac{dE}{dt} = \frac{1}{5} \big \langle \dddot{I}_{jk} \dddot{I}^{jk} \big \rangle
\end{equation}
Here $I_{jk}$ is the \emph{reduced quadupole moment} of the mass distribution and $\dddot{I}_{jk}$ is its triple derivative with respect to time.

However, a true BH--BH coalescence distorts spacetime in a highly non-linear manner, so we are forced to solve the exact Einstein's equations rather than their linear approximations.  In addition, since gravitational radiation is only well defined at spatial infinity, information from the waves must be extracted in a region of spacetime that is as far from the interaction region as we can practically get (where the background is as flat as possible).

There is an extensive literature surrounding this subject, including a recent historical survey\cite{cervantes} and a number of technical references \cite{flanhughes, baumshap4, alcubierre6, bishop, moncrief, weyltensor}, but the basics are as follows: using the so-called ``transverse-traceless" (TT) gauge,\cite{baumshap5} $h_{\mu\nu}$ is decomposed into the two possible polarization states for gravitational waves,\cite{baumshap6} $h_{+}$ and $h_{\times}$.  Note that the ``plus" and ``cross" polarization axes are rotated by $45^{\circ}$ with respect to each other.  In the Newman-Penrose treatment\cite{weyltensor} of gravitational radiation, which is used in most of the recent numerical gravitational-wave calculations, 
 the Weyl scalar $\Psi_4$ can (to linear order) be associated with outgoing gravitational radiation at spatial infinity.   One can then find its multipole content by using spin-2 weighted spherical harmonics to perform the orthogonal decompositions of the calculated waves (\emph{e.g.} Fig.\,\ref{nrcomparison}).   Recall that the lowest-order term in the multipole expansion of a gravitational wave is the quadrupole ($l=2$) with five $m$-state projections $(-2\le m \le +2)$.  Higher-order terms (so-called sub-dominant modes) can also appear\cite{hom} but in general are weaker.

\subsection{Numerical calculations of BH--BH coalescence}\label{numcalc}
The pioneering Hahn--Lindquist\cite{hahnlind} computation treated two equal-mass BHs colliding head-on (See Sec.\,\ref{exmp}).  
A dozen years later, Smarr and collaborators\cite{smarr1} used a similar model to study the head-on collision of non-rotating BH's with emission of gravitational radiation.  

While neither of these calculations converged to a physically reasonable result, at the time there appeared to be no fundamental obstacle to achieving realistic results once enough computational power could be brought to bear.  The stability issues mentioned in Sec.\,\ref{evcon}, especially regarding hyperbolicity, maintaining constraints, and how best to handle the physical BH singularities were not yet fully appreciated.  In addition, a full 3-D calculation of a general BH--BH coalescence proved to be far more difficult than expected.  Dealing with these issues awaited the arrival of black hole excision (1987), BSSNOK (1987 to 1998), GHCD (2005),  and the ``moving punctures'' (2005) algorithms.  Overall, these developments conservatively required over 40 years of effort.

\begin{figure}[h]
\centering
\includegraphics[width=8.5 cm]{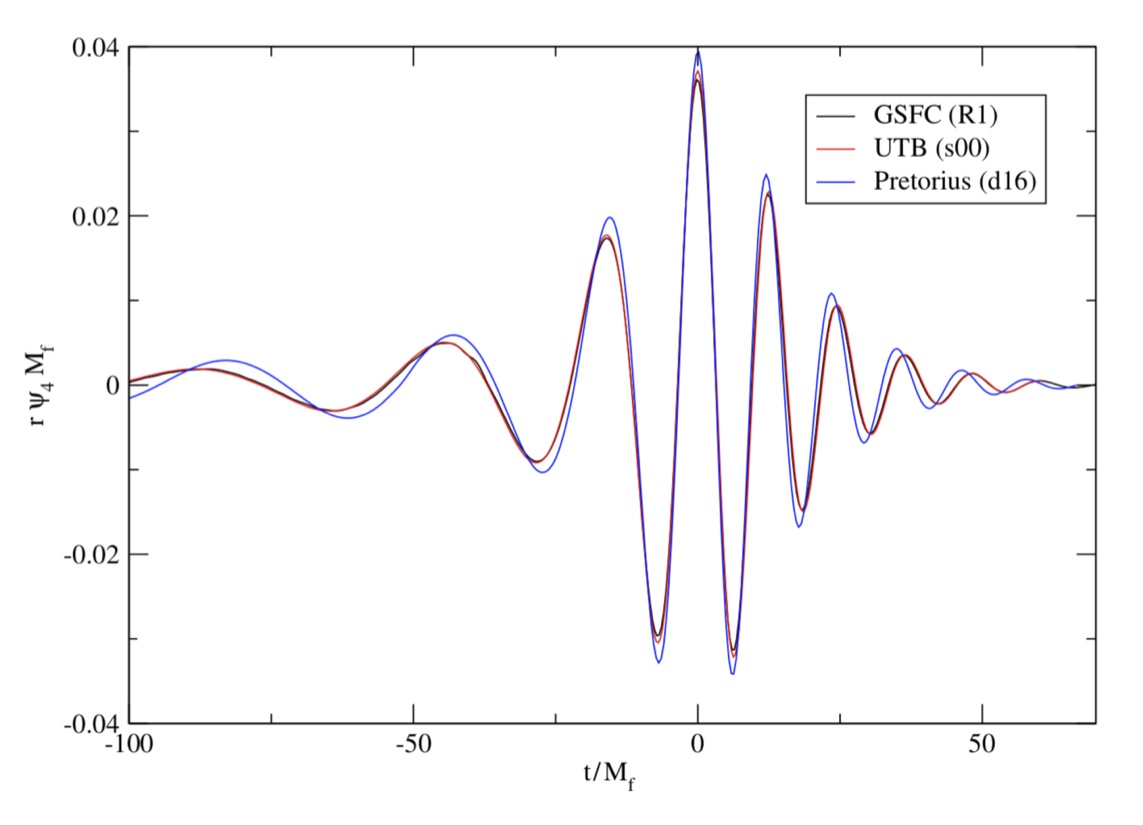}
\caption{Comparison of calculations from Pretorius\cite{pretorius} (red), Campanelli \emph{et al.}\cite{brownsville}  (blue) and Centrella \emph{et al.}\cite{goddard} (black).  The abscissa shows time (in units of the final BH mass) and the ordinate is the + polarization (the real part of the $l=2, m=2$ component) of the outgoing gravitational radiation. 
(From Ref.\,[\citenum{Baker:2007fb}])}
\label{nrcomparison}
\end{figure}

In 2005, great breakthroughs were achieved by Pretorius\cite{pretorius} and the Brownsville\cite{brownsville} and Goddard\cite{goddard} groups.  Working independently and using quite different methods, they performed stable, accurate simulations of BH--BH coalescence that agreed very well with each other.\,\cite{Baker:2007fb}  Fig.\,\ref{nrcomparison} compares their calculations of the real part of the $l=2, m=2$ mode (the $+$ polarization) of the gravitational waveform for a head-on collision of equal-mass BHs resulting in the formation of a Kerr BH.  Pretorius\cite{pretorius} used the GHCD formulation and BH excision.  The Brownsville\cite{brownsville} and Goddard\cite{goddard} groups used the BSSNOK formulation with the BHs represented by moving punctures.  These early calculations all employed FD integration methods.

It is not possible to overstate the importance of these results.  With reliable, highly accurate numerical methods in hand, not only is the full scientific content of the gravity-wave detections revealed, but more realistic calculations are possible (\emph{e.g.} that can include many orbits, unequal BH masses, and the effects of spin on the orbital motion).  Detailed calculations of more complicated gravitational systems, such as NS binaries\cite{nsns, bnsevol} or NS--BH systems\cite{nsbhcoalesc}, as well as detailed tests of strong-field general relativity\cite{testsofgr150914}, have begun.  Many codes have been developed;\nolinebreak
\footnote{\,\,See Refs. [\citenum{nrcodes}] and [\citenum{ninjaproject}] for comparative discussions.}
most use a BSSNOK+FD framework, the others a GHCD+SpEC treatment.

\subsection{Inspiral -- Merger -- Ringdown (IMR) models}\label{imr}
To identify possible BH--BH mergers and obtain estimates of their physical parameters, the data analyses use ``template banks'' of strain waveforms that can be matched nearly in real time with incoming strain data.

However, assembling a template bank is a major challenge.  Because templates can depend on as many as 17 parameters, thousands to millions of them are needed to span the parameter space.  Since each  fully relativistic calculation takes weeks to months to do, this is a totally impractical goal.  In addition, such calculations become prohibitively expensive as the number of orbits increases (beyond $\sim$ 20), when the BH-BH mass ratio is large (beyond $\sim 10$), or when the (normalized) BH spins are near unity.   For these reasons, the existing publicly available fully relativistic waveform catalogs\nolinebreak
\footnote{\,\,These include NINJA\cite{ninjaproject, ninja1, ninja2, ninja2add} (56 simulations), NRAR\cite{nrar} (25), Georgia Tech\cite{gatech} (452), RIT\cite{rit} (126), sXs\cite{sxs} (1425, 316 of which are public) and LVC\cite{lvcsims} (340).} 
contain in sum at most 2,500 templates.\cite{nrgwo}


This difficulty has led to the development of highly efficient hybrid models that capitalize on the fact that the great majority of the BH--BH coalescence waveform can be described by calculations that do not require numerical solutions to strong-field general relativity.  As mentioned in Sec.\,[\ref{overview}], analytical EOB and PN methods give accurate accounts of the BH-BH inspiral up to just before merger, and quasi-normal mode (QNM) analytical descriptions give accurate representations of the post-merger ringdown of the resulting Kerr BH. What is missing is an accurate representation of the strong non-linear fields in the region of the merger.   These are accounted for by calibrating the hybrid models against waveforms calculated using full numerical relativity.  The calibrations effectively provide a phenomenological representation of the merger and post-merger waveform that interpolates between, and extends beyond, the available NR simulations.  But even without this calibration, the EOB--QNM models are qualitatively, and semi-quantitatively, correct.  

Three recent examples of how this is done are the \texttt{SEOBNRv4}\cite{seobnrv4}, \texttt{TEOBResumS}\cite{teobresums} and \texttt{IMRPhenomPv3}\cite{imrphenompv3} simulations.  \texttt{SEOBNRv4} and \texttt{TEOBResumS} are based on time-domain EOB formalisms that can describe the coalescence of spinning, non-precessing BBHs through ringdown, and for binary neutron stars up to merger.   \texttt{IMRPhenomPv3} can describe precessing BBHs incorporating two-spin effects.  The model is based in the frequency domain, resulting in a very much faster execution time.  It is validated against a set of precessing numerical relativity simulations.  

Recently, \emph{surrogate models} have been developed that are based on interpolation in the parameter space of existing fully relativistic calculations.  A recent version, NRSur7dq2,\cite{surrogatemod} includes all seven dimensions of the parameter space (\emph{i.e.} mass ratio and the spins).

\subsection{Some future pathways for numerical relativity and gravitational wave science}
In the short period since September 14, 2015, the discovery and interpretation of gravitational waves has shown itself to be a revolutionary new means of studying the Universe.  This is true even though the volume reach of current detectors is but a very small part 
of what it could be with improved (but realistic) detectors of enhanced sensitivity.  It is reasonable to expect that a network of such \emph{Third Generation (3G)} detectors would be able to study phenomena occurring at the edge of the observable Universe.  Events taking place closer to home would be seen in far greater numbers than at present, and with much higher signal-to-noise ratios.  Thus it is not surprising that there is intense interest in developing the science case for such an enhanced program.  The \emph{Gravitational Wave International Committee, GWIC},\nolinebreak
\footnote{\,\,GWIC is comprised of representatives from the world's gravitational wave observatories. Its purpose is to promote the field via international planning and cooperation.\cite{gwic}}
is taking the lead role in conceptualizing how such a future might be realized.

A recent very extensive ``roadmap"\cite{barack} catalogues many intriguing opportunities on the horizon.  Since numerical relativity has played such an essential role in the development of  gravitational wave science to date, it is clear that it will play an indispensable part in identifying realistic 3G  science goals.

Possibilities for exploration at some heretofore unapproachable frontiers of cosmology are (in the near term, before the advent of a 3G network):
\vspace{2mm}
\begin{itemize}
\item Further tests of strong-field general relativity.  (See Refs.\,[\citenum{testsofgr150914, yunesyagipre, testsofgr170817}] for current results);
\item Decoding the structure and dynamics of BHs and their population distributions\cite{imbhsearch, curiel};
\item Decoding the structure and dynamics (the equation of state) of neutron stars and their population distributions\cite{nsnsc, nsradii, metzger};
\item Possible detection of gravitational waves from core-collapse supernovae\cite{gossan};
\item Constraints on evidence for cosmic strings\cite{cosmstrgs};
\item Possible detection of polarizations in the stochastic gravity wave background\cite{polar, colacino};
\item Other prospects in astrophysics\cite{astrophyimps, lehpre, garf2, hubblecon}.
\end {itemize}
\vspace{2mm}

Getting the most out of these studies will require advances in these additional areas (among others):
\vspace{2mm}
\begin{itemize}
\item Improving numerical algorithms\cite{senr1, seobnrv4, imrphenompv3, surrogatemod} for much more efficient creation of template banks;
\item Extending template banks to include more orbits, larger mass ratios, and higher spin values in BBH coalescences\cite{nrar, gatech, rit, sxs, lvcsims};
\item Further study of ways to determine BH spins and the roles of spin orientations and orbit precession in BBH coalescences\cite{hannampreces, constbhspin}; 
\item Achieving a better understanding of ambiguities can that occur in parameter estimation\cite{constbhspin, mrsdegen, spindegen}
\item Developing strategies for handling possibly large rates of overlapping events that may appear when more sensitive detectors are online \cite{regimbau}; and
\item Improving methods for more robust multi-messenger astronomy collaborations.\cite{mmacollab}
\end {itemize}
\vspace{1mm}
These lists are themselves dynamic and will certainly change, perhaps radically, as the field develops further.  This is testament to a field full of promise.

\subsection{You can try this at home}
Should you wish to do calculations on your own, there are very helpful resources available: consult the \emph{Simulating Extreme Spacetimes (SXS)},\cite{sxs}  \emph{Einstein Toolkit}\cite{etk} and \emph{Super Efficient Numerical Relativity (SENR)}\cite{senr1, senr2} websites for more information.  Refs.\,\citenum{baumshap1, alcubierre1, gour1} 
also offer numerical examples.  The LIGO Open Science Center\cite{losc} provides data from gravitational-wave observations along with access to tutorials and software tools.   You can also participate in the LIGO search for gravitational waves  by signing up with Einstein@Home.\cite{einstathome}  

\section{Final Comment}
GW150914 was a figurative supernova in the history of physics and cosmology.  It, and the LIGO/VIRGO discoveries since then, have amazed even the most optimistic among us.  GW170817, the first-ever sighting of a NS--NS merger and its subsequent electromagnetic counterparts, has provided a remarkable glimpse of the power of multi-messenger astronomy.  The last three years have revealed just how much the ``gravitational Universe'' has to teach us now that we can see it.

It has taken 100 years to reach this point.  Because of the genius of Albert Einstein, who saw that the geometry of the Universe was more subtle than realized by Isaac Newton, and  the incredible ingenuity of the students, engineers and scientists of the gravitational science community, we can now use gravitational waves as a tool to decode the Universe.  But without the generosity and patience of our fellow citizen-scientists the world over, these discoveries would not have been possible.

\section{Acknowledgments}
This paper is an extended version of an introductory presentation I made on this subject at MIT-LIGO in December, 2016.
I gratefully thank my MIT colleagues for many conversations about gravitation and cosmology, as well as  Bruce Allen, Thomas Baumgarte, Manuela Campanelli, Matt Evans, Evan Hall, Mark Hannam, Erik Katsavounidis, Rob Owen, Harald Pfeiffer, Mark Scheel, Frank Tabakin and Rai Weiss for very useful contributions to this manuscript.


\begin{thebibliography}{000}
\bibitem{lapseshift1} For a concise discussion see Ref.\,[\citenum{centrev}], p. 13 or Ref.\,[\citenum{alcubierre5}], p. 30.   
More extensive treatments are found in Ref.\,[\citenum{baumshap1}], Ch. 4 and Ref.\,[\citenum{alcubierre1}],  Ch. 4.

\othercit
\end{thebibliography}

\end{document}